\documentclass[aip,numerical,reprint,jmp,showpacs,showkeys,
  onecolumn]{revtex4-1}

\usepackage{amsmath,amsthm,amssymb,amscd,fancybox,ifthen,float,subfigure}
\floatstyle{plain}

\usepackage{graphicx}
\usepackage[all]{xy}
\usepackage[margin=1in]{geometry}

\theoremstyle{definition}

\theoremstyle{remark}

\usepackage{cmbright}

\floatstyle{plain}
\restylefloat{figure}

\newcommand\fr{\widehat r}
\newcommand\fq{\widehat q}
\newcommand\fg{\widehat g}
\newcommand\bj{\boldsymbol j}
\newcommand\bmm{\boldsymbol m}
\newcommand\bG{\boldsymbol G}
\newcommand\bI{\boldsymbol I}
\newcommand\bn{\boldsymbol n}
\newcommand\br{\boldsymbol r}
\newcommand\bx{\boldsymbol x}
\newcommand\by{\boldsymbol y}
\newcommand\bA{\boldsymbol A}
\newcommand\bD{\boldsymbol D}
\newcommand\bQ{\boldsymbol Q}
\newcommand\bHtot{\boldsymbol H^\text{tot}}
\newcommand\bE{\boldsymbol E}
\newcommand\bH{\boldsymbol H}
\newcommand\bEtot{\boldsymbol E^\text{tot}}
\newcommand\bEin{\boldsymbol E^\text{in}}
\newcommand\bHin{\boldsymbol H^\text{in}}
\newcommand\bJ{\boldsymbol J}

\newcommand\bz{\hat{\boldsymbol{z}}}
\newcommand\bbR{\mathbb R}

\begin{document}

\title{A generalized Debye source approach to electromagnetic
  scattering in layered media}

\date{\today}

\author{Michael O'Neil}
\email[Email: ]{oneil@cims.nyu.edu}
\noaffiliation
\affiliation{Courant Institute, New York University}

\begin{abstract}
  The standard solution to time-harmonic electromagnetic scattering
  problems in homogeneous layered media relies on the use of the
  electric field dyadic Green's function. However, for small values of
  the governing angular frequency~$\omega$, evaluation of the electric
  field using this Green's function exhibits numerical instability.
  In this short note, we provide an alternative approach which is
  immune from this low-frequency breakdown as $\omega \to 0$. Our
  approach is based on the generalized Debye source representation of
  Maxwell fields.  Using this formulation, the electric and magnetic
  fields gracefully decouple in the static limit, a behavior similar
  to that of the classical Lorenz-Debye-Mie representation of Maxwell
  fields in spherical geometries. We derive extensions of both the
  generalized Deybe source and Lorenz-Debye-Mie representations to
  planar geometries, as well as provide equations for the solution of
  scattering from a perfectly conducting half-space and in layered
  media using a Sommerfeld-like approach. These formulas are stable as
  $\omega$ tends to zero, and offer alternatives to the electric field
  dyadic Green's function.
\end{abstract}



\maketitle

\section{Introduction}

Scattering in a half-space and layered media, both acoustic and
electromagnetic, has been a classic problem in the physics of waves
for many decades. Solutions for various boundary conditions have been
formulated by Van der Pol\cite{vanderpol},
Sommerfeld\cite{sommerfeld}, and Weyl\cite{weyl}. Generalizations of
this problem are of special interest in non-destructive testing of
microchip fabrication, oil speculation, geological surveying,
materials science, etc.  There have been several methods and
formulations developed to address the construction of the
corresponding Green's
function\cite{lindell1,lindell2,lindell3,hochmann,taraldsen,oh-1994}
and embed the resulting analysis into a fast numerical algorithm that
can be used industrially\cite{geng-2000}. Approaches which are based
on the electric field dyadic Green's function, however, commonly
suffer from numerical instabilities in the static limit as the
governing frequency $\omega$ tends to zero.  This Green's function
includes a term of $\mathcal O(\omega^{-2})$, which has to be
necessarily offset by catastrophic numerical cancellation. The main
purpose of this note is to present an alternative approach to the
problem of electromagnetic scattering in layered media which does {\em
  not} suffer from numerical instabilities as $\omega \to 0$. Our
approach is based on the generalized Debye source
representation\cite{EpGr,EpGrOn,EpGrOn2} of Maxwell fields. Using this
representation, as $\omega \to 0$, the electric field and the magnetic
field gracefully and stably uncouple to their respective static
limits. Equations for the scattered electric field only depend upon
the impinging electric field, and vice versa for the magnetic field.
This is analogous to the behavior exhibited by the classic
Lorenz-Debye-Mie representation of Maxwell fields in the exterior of a
sphere\cite{debye-1909,mie-1908,bouwkamp,wilcox}. Along the way, as
motivation for the generalized Debye source approach, we will also
reformulate the Mie series solution for spherical scattering as one
which is based on plane waves and is compatible with planar Cartesian
geometries.

In focusing on electromagnetic waves in linear,
isotropic, non-dispersive, planar layered domains (and half-spaces),
we will restrict ourselves to the time-harmonic case, assuming a
dependence of $e^{-i\omega t}$ which will be suppressed from now on.
Under these assumptions, the fully time-dependent Maxwell's equations
reduce to the set of equations:
\begin{equation}
\begin{aligned}
\nabla \times \bE &= i \omega \mu \bH, & \qquad
 \nabla \times \bH &= -i \omega \epsilon \bE + \bJ , \\
\nabla \cdot \bE &=\rho/\epsilon,  & \nabla \cdot  \bH &= 0,
\end{aligned}
\end{equation}
where $\omega$ is the angular frequency, $\mu$ is the magnetic
permeability, and $\epsilon$ is the electric permittivity
\cite{jackson}. Furthermore, the physical electric current $\bJ$ and
electric charge $\rho$ must also satisfy the continuity
condition
\begin{equation}
\nabla \cdot \bJ = i \omega \rho.
\end{equation}
The Helmholtz parameter (wavenumber) $k=\omega \sqrt{\epsilon\mu}$
will be assumed to have positive real part and non-negative imaginary
part in order to ensure causality. This assumption on $k$ also causes
all fields constructed from layer potentials using the Helmholtz
Green's function to automatically satisfy
the Silver-M\"uller decay condition at infinity:
\begin{equation}
  \lim_{|\bx| \to \infty} \left( \bH(\bx) \times \hat\br -
  \frac{\mu}{\epsilon} \bE(\bx) \right) = 0.
\end{equation}

In this note, we will mainly be concerned with
solving Maxwell's equations in the context of scattering phenomena. In
particular, we will write the {\em total} fields $\bE^\text{tot}$,
$\bH^\text{tot}$ as the sum of a known incoming field and an unknown
scattered field, i.e.
\begin{equation}
\bE^\text{tot} = \bE^\text{in} +\bE, \qquad
\bH^\text{tot} = \bH^\text{in} +\bH.
\end{equation}

Historically, most of the effort in constructing solutions to Maxwell's
equations in a half-space and layered media has been focused on
constructing the associated Helmholtz (arising in transverse
electromagnetic problems) or Maxwell dyadic Green's
function\cite{chew,cai,lindell1,lindell2,lindell3,michalski-1997} (as
described in Section~\ref{sec-dyadic}).
These approaches, while powerful and readily adaptable to different
geometries, can suffer from numerical instabilities for small
$\omega$, and often require complicated quadrature schemes to handle a
variety of non-generic singularities in the resulting integral
representations (as can be the case in the method of complex
images\cite{taraldsen,oneil-imped,thomson-1975}).

In the following discussion, we propose an alternative approach to the
problem of electromagnetic scattering in layered media which does not
suffer from numerical instabilities as $\omega \to 0$.  To this end,
our main result is a Sommerfeld-like (spectral) formulation of the
generalized Debye source representation of Maxwell fields which is
well-conditioned for any value of $\omega$, including $\omega=0$.
This behavior is consistent with that of the generalized Debye source
formalism in the case of scattering from arbitrary smooth bounded
obstacles, unlike most representations based on the electric field
dyadic Green's function.  In the presence of a perfectly conducting
half-space, our new representation completely decouples at {\em any}
frequency $\omega$, not only in the static case.  In layered media
geometries, there is an inter-layer coupling of unknowns which becomes
weaker as $\omega \to 0$.  As a preliminary warm-up, we first extend
the Lorenz-Debye-Mie construction of Maxwell fields to Cartesian
planar domains.  This derivation, while straight-forward and a natural
limit of the spherical case, seems to be absent from the scattering
literature.  In the case of the perfectly conducting half-space, the
Debye potentials are coupled via a two-by-two linear system. This
extends to the layered media case with an inter-layer coupling of
unknowns, with four Debye potentials defined on each interface.  As in
the usual spherical Mie representation, the fields decouple as $\omega
\to 0$, resulting in formulas which are numerically stable.

The paper is organized as follows: In Section~\ref{sec-theproblem} we
describe the half-space and layered media electromagnetic scattering
problems and introduce notation that will be used in the rest of the
discussion. Section~\ref{sec-dyadic} briefly reviews some of the most
common methods of constructing half-plane and layered media solutions,
those based on the dyadic Green's function for Maxwell's
equations. Next, in Section~\ref{sec-mie} we introduce the first of
our new representations for layered media scattering by extending the
classic Lorenz-Mie-Debye spherical representation of solutions to a
half-space.  The main contribution of this note is in
Section~\ref{sec-debye}, where we extend the generalized Debye source
representation to an infinite half-space and layered media. Both new
representations in Sections~\ref{sec-mie} and~\ref{sec-debye} provide
formulas which are numerically stable in the low-frequency limit.
Section~\ref{sec-conclusions} contains closing remarks on the methods
previously introduced.

Lastly, the appendix contains useful Fourier transform identities for
(spherical) partial wave functions that can be used to analytically
represent incoming fields in Cartesian coordinates. Using these
formulae, an algorithm for scattering in layered media, analogous to
Mie scattering, can be derived.

\section{Half-space and layered media scattering problems}
\label{sec-theproblem}

\begin{figure}[t]
  \subfigure[Perfect conductor.]{\raisebox{.4\height}
    {\includegraphics[width=.3\linewidth]{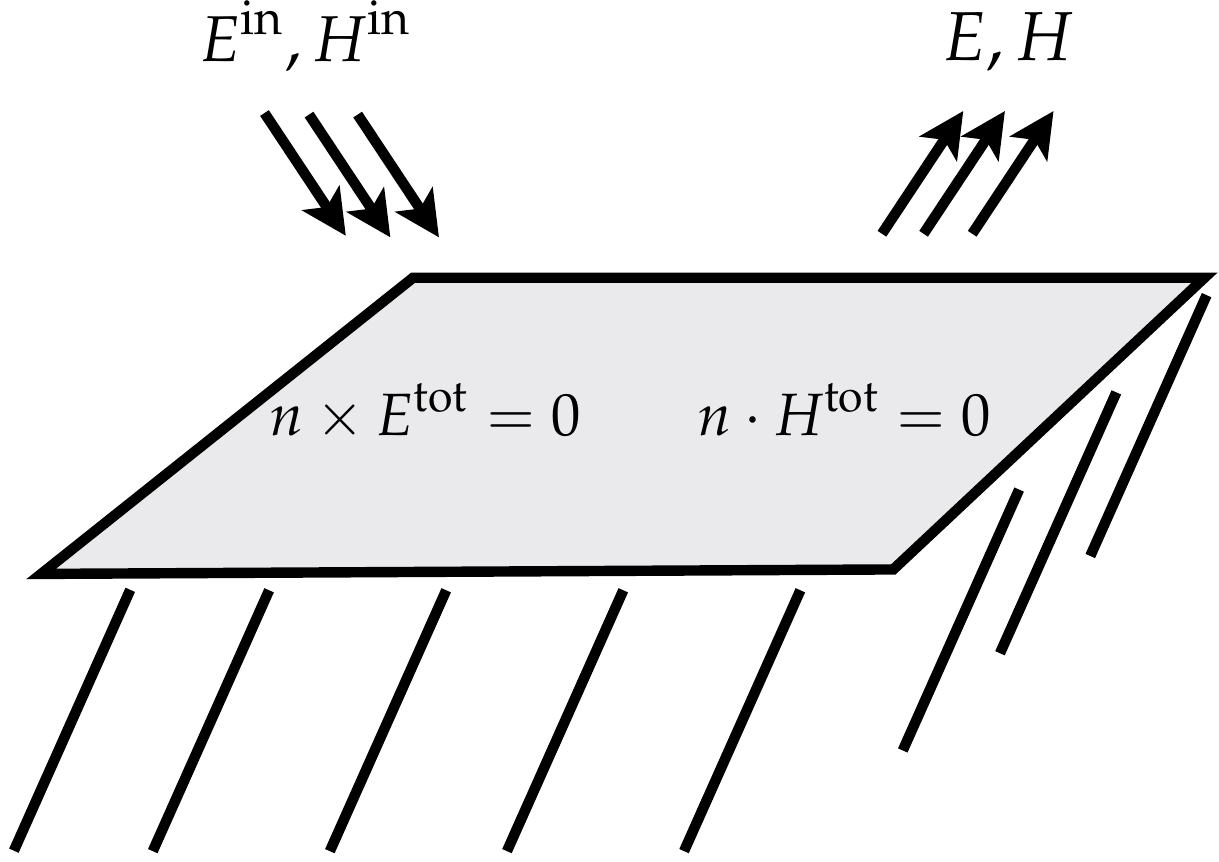}}} \quad
  \subfigure[Bi-layered media.]{\raisebox{.65\height}
    {\includegraphics[width=.3\linewidth]{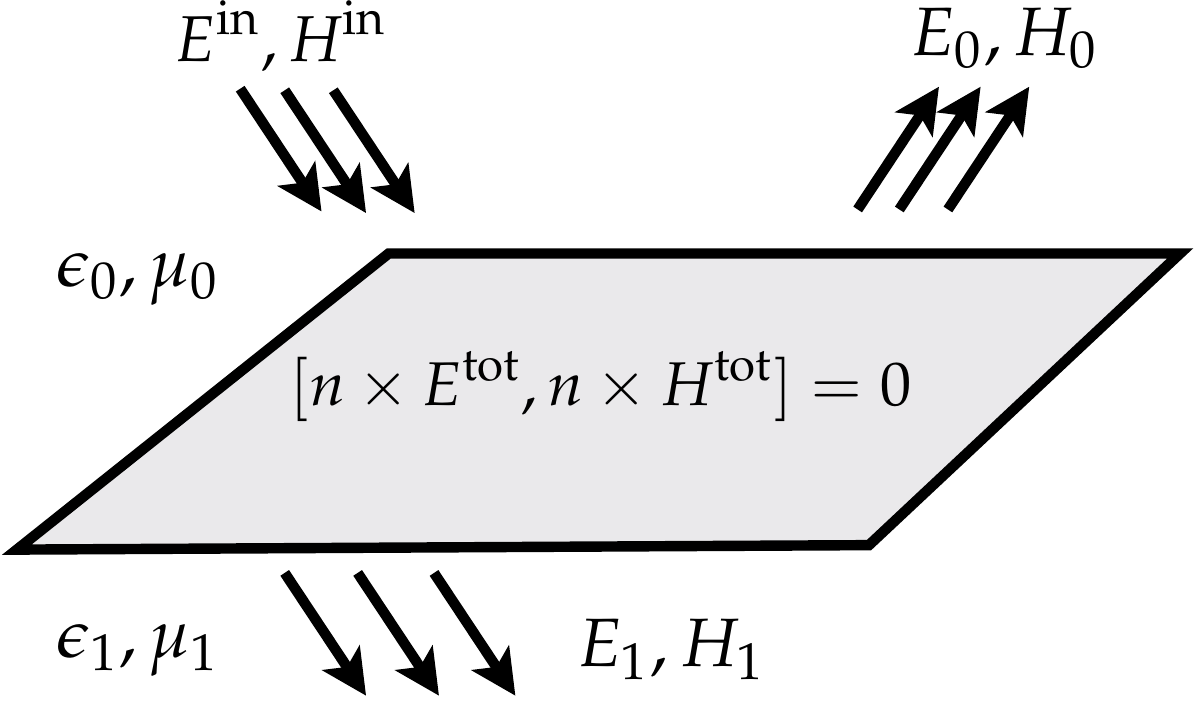}}} \quad
  \subfigure[Multi-layered media.]{\raisebox{0\height}
    {\includegraphics[width=.3\linewidth]{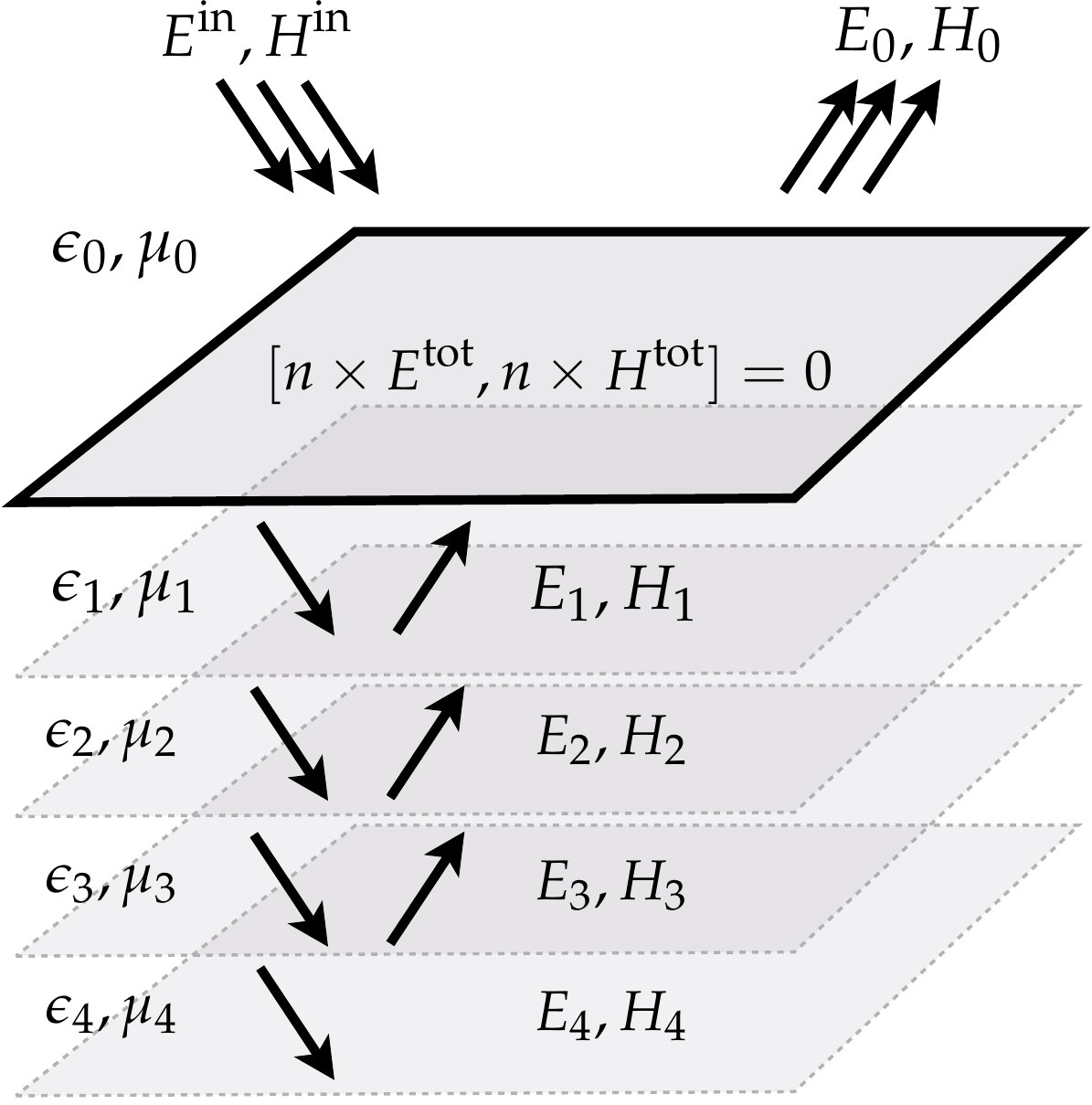}}}
  \caption{A graphical depiction of the three main categories of planar
    scattering.}
  \label{fig-3probs}
\end{figure}

Half-space and layered media electromagnetic scattering problems
largely fall into three main categories: perfectly-conducting or
impedance half-space problems, homogeneous bi-layer transmission
problems, and multiple-layer transmission problems. See
Figure~\ref{fig-3probs} for a graphical depiction of the geometry
inherent in each class of problems. We will always assume that the
region $z>0$ is homogeneous with material parameters $\epsilon$, $\mu$
or ($\epsilon_0$, $\mu_0$ in the case of multiple layers).

In the case of a perfectly conducting half-space $z\leq 0$, the four
physical boundary conditions on the plane $z=0$ are:
\begin{equation}
\begin{aligned}
\bn \times \bE^\text{tot} &= 0, & \qquad \bn \cdot \bEtot &= \rho, \\
\bn \times \bH^\text{tot} &= \bJ, & \bn \cdot \bHtot &= 0,
\end{aligned}
\end{equation}
where $\bn = \bz$ (the unit-normal in the $z$-direction) and $\bJ$,
$\rho$ are the physical current and charge, respectively. Depending on
the representation of $\bE$, $\bH$, one or more of the above boundary
conditions may be enforced, or a complex linear combination of
multiple conditions may be used.  Using the standard
representation of $\bE$ and $\bH$ in the Lorenz gauge,
\begin{equation}\label{eq-lorenz}
\begin{aligned}
\bE(\bx) &= i\omega\mu \int g_k(\bx,\bx') \, \bJ(\bx') \, dV'
 - \frac{1}{i\omega\epsilon}
\nabla \int g_k(\bx,\bx') \, \nabla \cdot  \bJ(\bx') \, dV', \\
\bH(\bx) &= \nabla \times \int g_k(\bx,\bx') \, \bJ(\bx') \, dV',
\end{aligned}
\end{equation}
conditions on the tangential fields lead to the electric field
integral equation (EFIE) or the magnetic field integral equation
(MFIE).  Above, $g_k$ is the Green's function for the
three-dimensional Helmholtz equation with parameter $k$.  The EFIE is
a hypersingular integral equation for the current $\bJ$ which can be
regularized\cite{contopanagos-2002} via Calderon projections, but
still suffers from low-frequency breakdown.  The MFIE is stable on
simply-connected geometries as $\omega \to 0$\cite{epstein-2013}, but
for sufficiently small $\omega$ the electric field cannot be recovered
without solving an additional integral
equation\cite{vico-2013}. Complex linear combinations of the EFIE and
the MFIE yield the combined-field integral equation (CFIE), which is
free from spurious resonances in $k$, however still susceptible to
instabilities for small $\omega$\cite{song-1995}.  It is assumed that
the scattered field from a perfectly conducting half-space 
adheres to the usual Silver-M\"uller radiation condition - this can be
seen from an analysis of the domain Green's function (which is
analytically given in Section~\ref{sec-dyadic}).

In the case of layered dielectric materials, the physical transmission
boundary conditions between layers can be phrased in terms of
continuity in the tangential components of $\bEtot$, $\bHtot$, or in
the normal components or normal derivatives of $\epsilon\bEtot$,
$\mu\bHtot$:
\begin{equation} \label{eq-dibc}
  \begin{aligned}[]
    [\bn \times \bE^\text{tot}] &= 0, & \qquad [\bn \cdot
    \epsilon\bEtot] &= 0, & \qquad \left[ \bn \cdot
      \epsilon\frac{\partial\bEtot}{\partial n}
    \right] &= 0, \\
    [\bn \times \bH^\text{tot}] &= 0, & [\bn \cdot \mu\bHtot] &= 0, &
    \left[ \bn \cdot \mu\frac{\partial\bHtot}{\partial n} \right] &=
    0.
\end{aligned}
\end{equation}
The notation $[ f ]$ is used to denote the discontinuity in $f$ across
the boundary. We use this notation as it is convenient for multiple
layers. The usual formulation of dielectric problems under these
boundary conditions is due to M\"uller\cite{muller}, which enforces
the tangential boundary conditions. This formulation, however, still
suffers from numerical instabilities as $\omega \to 0$ due to the
integral representation that is used\cite{EpGrOn}.  In layered media
problems, the radiation condition imposed on the scattered field is
somewhat different than that in free-space, the Silver-M\"uller
condition. The proper radiation condition can be derived from an
asymptotic analysis of the domain Green's
function\cite{duran-2009}. It suffices to point out that if the
dielectrics are slightly absorbing, i.e. have a wavenumber $k_j$ with
small imaginary part, then non-decaying surface waves are prohibited.
Integral equation formulations of the two previous problems on bounded
domains have been updated using generalized Debye source
representations, which are immune from low-frequency breakdown,
topological breakdown (in multiply connected geometries), and
spurious resonances in $k$\cite{EpGr,EpGrOn}.

So far, we have ignored the topic of discretization. Generally, it is
not feasible to use the physical variables $\bJ$ and $\rho$ to
discretize the infinite interfaces between homogeneous dielectrics, a
prohibitively large linear system would be the result. Therefore,
there are two main approaches to the problem: construct the domain
Green's function which accounts for all the boundary conditions
between layers, or discretize the Fourier transform of the unknowns
defined on the interfaces, which will be numerically compactly
supported if the original unknown is smooth. This, latter, spectral 
approach was introduced by Sommerfeld\cite{sommerfeld}.

In the case of the perfectly conducting half-space or bi-layer
dielectric, the dyadic Green's function can be readily
constructed\cite{cai,lindell1,lindell2}, although possibly taking on a
complicated analytical or integral form.  In the presence of multiple
dielectric layers, no analytic closed-form solution for the Green's
function exists. In fact, the true Green's function corresponds to the
Green's function for a one-dimensional wave equation with non-constant
(piecewise-constant in our case) coefficients\cite{chew}.
Constructing an analytical approximation or a convergent numerical
scheme has been the subject of countless classical electromagnetics
and electrical engineering papers.

It should be mentioned that one analytical solution does exist, that
of scattering in layered media by pure plane waves (transverse
waves). The evaluation of the scattered field can be reduced to
recursively calculating reflection and transmission coefficients layer
by layer\cite{chew}. This approach can handle several real-world
scattering problems, but is not suitable for arbitrary incoming fields
or being embedded inside simulations involving complicated geometry.
For example, if the incoming field is generated by dipoles, the
spectral components of the field must first be calculated in order for
the method of reflection/transmission of plane waves to be applied.

On the other hand, Fourier methods, commonly referred to as Sommerfeld
methods, are applicable in the presence of arbitrarily smooth incoming
fields.  These methods hinge on the Fourier transform of the Helmholtz
Green's function, which is given by:
\begin{equation}
\frac{1}{8\pi^3} \int_{\bbR^3} \frac{e^{ik|\bx|}}{4\pi |\bx|}
\, e^{-i(\xi x + \eta y + \lambda z)} \, dx dy dz
= \frac{1}{8\pi^3} \frac{1}{\xi^2 + \eta^2 + \lambda^2 - k^2}.
\end{equation}
If $g_k$ is written in terms of it's Fourier transform, and
the $\lambda$ integral is evaluated via contour integration\cite{chew},
then we have the following Sommerfeld formula for $g_k$:
\begin{equation}\label{eq-som}
\begin{aligned}
g_k(\bx,\bx') &= \frac{e^{ik|\bx-\bx'|}}{4\pi |\bx-\bx'|} \\
&= \frac{1}{8\pi^3} \int_{\mathbb R^3} 
\frac{e^{i(\xi(x-x') + \eta(y-y') + \lambda(z-z'))}}
{{\xi^2 + \eta^2 + \lambda^2 - k^2}} \, d\xi d\eta d\lambda \\
&= \frac{1}{8\pi^2} \int_{\mathbb R^2} 
         \frac{e^{-\sqrt{\xi^2+\eta^2-k^2}|z-z'|}}{\sqrt{\xi^2+\eta^2-k^2}}
         e^{i(\xi(x-x') + \eta(y-y'))} \, d\xi \, d\eta.
\end{aligned}
\end{equation}
In the following sections, we will denote by $\widehat f = \mathcal F[f]$ the
two-dimensional Fourier transform of a function $f$. Using this
notation, $\widehat g_k$ is given by:
\begin{equation}
\begin{aligned}
\widehat g_k(\xi,\eta,z) &= \mathcal F[g_k] \\
&= \frac{1}{8\pi^2} 
\frac{e^{-\sqrt{\xi^2+\eta^2-k^2}|z|}}{\sqrt{\xi^2+\eta^2-k^2}}.
\end{aligned}
\end{equation}
The methods described in
Sections~\ref{sec-mie} and~\ref{sec-debye} do not construct the domain
Green's function directly, but rather rely on Sommerfeld
representations of the Helmholtz Green's function. Unfortunately,
these methods result in slow numerical convergence when the incoming
field is generated by a scatterer which is near an interface (due to a
slowly converging Fourier transform density, similar to the one in the
above formula). Schemes that combine the benefits of Green's function
methods (images) with the rapid convergence of far-field Sommerfeld
contributions have been recently developed\cite{oneil-imped}.

\section{The dyadic Green's functions}
\label{sec-dyadic}

One of the most popular tools used in numerical simulations of
electromagnetic fields in layered media is the dyadic Green's
function. Evaluations of the dyadic Green's function due to Hertz
current dipole sources (either in real-space or Fourier-space)
oriented horizontally and vertically can be linearly combined to
construct the response for arbitrarily oriented current
sources\cite{cai,chew}. These techniques have been at the center of a
large number of layered media scattering schemes, and are widely
applicable in complicated geometries because of the local natural of
the Green's function. Furthermore, they provide one possible solution
to the problem of scattering from objects which are arbitrarily close
to dielectric-dielectric or dielectric-perfectly conducting interfaces
as the induced field singularity can be explicitly handled using an
adaptive discretization scheme. However, the approach is not without
its drawbacks - namely, evaluation of electric fields via convolution
of the dyadic Green's function with a current source becomes
numerically unstable as $\omega \to 0$. Evaluation of the dyadic
Green's function requires differentiation followed by division by
$\omega^2$.

The nature of this low-frequency instability is the same as that which
is present in the standard representation of $\bE$ in the Lorenz
gauge, given in equation~(\ref{eq-lorenz}). Unless the divergence of
the electric current is handled analytically, with an {\em explicit}
factoring out of $\omega$, accuracy will be lost in the evaluation of
the scalar potential term since the resulting calculation suffers from
catastrophic cancellation since $\nabla \cdot \bJ \sim \mathcal
O(\omega)$. The spectral (Sommerfeld) form of the dyadic Green's
function suffers from the same form of numerical difficulties.
Methods which are based on so-called charge-current formulations are
one attempt to circumvent this
instability\cite{taskinen-2006,vico-2013,qian-2009}.

The dyadic Green's function for the electric field in
 Maxwell's equations \cite{vanbladel,chew} is given by
\begin{equation}
\begin{aligned}
\bar\bG^E(\bx,\bx') &= \left(\bar\bI + \frac{1}{k^2} \nabla\nabla
\right)\frac{e^{ik|\bx-\bx'|}}{4\pi |\bx-\bx'|} \\
&= \frac{1}{k^2} \nabla \times \nabla \times \left(
\frac{e^{ik|\bx-\bx'|}}{4\pi |\bx-\bx'|} \bar\bI \right),
\end{aligned}
\end{equation}
where a dyad $\bar\bD$ can be viewed as either a $3\times3$ matrix or
a rank-two tensor, and the matrix $\nabla\nabla$ has entries $\left(
\nabla \nabla \right)_{ij} = \partial^2/\partial x_i \partial x_j$.
Likewise, the dyadic Green's function for the magnetic field is:
\begin{equation}
\begin{aligned}
\bar\bG^H(\bx,\bx') &= \frac{1}{i\omega\mu} \nabla \times
\bar\bG^E_k(\bx,\bx') \\ &= \frac{1}{i\omega\mu} \nabla \times \bar\bI
\frac{e^{ik|\bx-\bx'|}}{4\pi |\bx-\bx'|}.
\end{aligned}
\end{equation}
For a localized distribution of electric current $\bJ$ in a
homogeneous region $V$, the induced electric and magnetic fields 
are given by
\begin{equation}
\begin{aligned}
\bE(\bx) &= i \omega \mu \int_V \bar\bG^E(\bx,\bx') \cdot \bJ(\bx')
\, dV_{x'}, \\
\bH(\bx) &= i \omega \mu \int_V \bar\bG^H(\bx,\bx') \cdot \bJ(\bx')
\, dV_{x'}. \\
\end{aligned}
\end{equation}
The Green's function for the perfectly conducting half-space can be
constructed explicitly via images\cite{jin-2010} in the lower
half-space,
\begin{equation}\label{eq-pecimages}
\bar\bG^E_{\text{pec}}(\bx,\bx') = 
\bar\bG^E(\bx,\bx') - \bar\bG^E(\bx,\bx_i') + 
2 \bar\bG^E(\bx,\bx_i') \cdot \bz \bz
\end{equation}
where the image points are given by $\bx_i' = (x,y,-z)$.  The images
in the above formula annihilate the tangential components of the
electric field on $z=0$.  In the presence of planar layered media,
the dyadic Green's function for the electric field must satisfy the
variable-coefficient vector wave equation:
\begin{equation}
\left( \nabla \times \frac{1}{\mu(z)} \nabla \times \bar\bI - \omega^2
\epsilon(z) \bar\bI \right) \bar\bG^E(\bx,\bx') = \frac{1}{\mu(z)} \bar\bI
\delta(\bx-\bx').
\end{equation}
A solution in the Fourier domain can be found to this equation using
vector wave functions (the vector version of partial wave expansions
for the Helmholtz equation, analogous to vector spherical harmonics),
but the result requires several calculations and would detract from
the following discussion.  An expression for the dyadic Green's
function using discrete real-images, like those in
equation~(\ref{eq-pecimages}) does not exist. See Section 7.4.2 in
Chew\cite{chew} for a thorough discussion of the above.
Unfortunately, both the physical and the spectral representation of
the dyadic Green's function suffers from low-frequency breakdown; both
include division by $k$ or $k^2$. Note that the magnetic field dyadic
Green's function does {\em not} suffer from low-frequency breakdown,
which is obvious from the behavior of the magnetic field integral
equation as $\omega \to 0$\cite{cools-2009} (however, other,
topological instabilities do arise).  The component-wise spectral
representation of $\bar\bG^E$ is given by:
\begin{equation}
\widehat{ \bar G_{jk}^E} = \frac{1}{8\pi^2}
\int_{\bbR^2} \left( \delta_{jk} + \frac{1}{k^2} 
\frac{\partial^2}{\partial x_j \partial x_k} 
\right)
\frac{e^{-\sqrt{\xi^2+\eta^2-k^2}|x_3-x_3'|}}{\sqrt{\xi^2+\eta^2-k^2}} \,
e^{i(\xi(x_1-x_1') + \eta(x_3-x_3'))} \, d\xi d\eta,
\end{equation}
where $(x,y,z) = (x_1, x_2, x_3)$ for the sake of convenient
notation. We now move onto the first of our new, stable
representations of electromagnetic fields in planar geometries, an
extension of the spherical Lorenz-Debye-Mie representation.

\section{Planar Mie scattering}
\label{sec-mie}

The analytical Mie series solution to scattering from perfectly
conducting and dielectric spheres is a classic result in
mathematical physics due to Mie in 1908\cite{mie-1908}, and then shortly
thereafter rederived by Debye while working on light pressure in
1909\cite{debye-1909}.  It is the Mie series solution that is often used to
determine whether time-harmonic or fully time-dependent
electromagnetic scattering codes are converging to the correct answer
at the correct rate\cite{chew}.

The spherical representation of such solutions is now referred to as
the Lorenz-Mie-Debye representation of time-harmonic electromagnetic
waves, and is given by:
\begin{equation}
\begin{aligned}
\bE &= \nabla \times \nabla \times  (u \, \hat\br) + 
i\omega\mu \nabla \times (v \, \hat\br), \\
\bH &= \nabla \times \nabla \times (v \, \hat\br ) - 
i\omega\epsilon \nabla \times ( u \, \hat\br),
\end{aligned}
\end{equation}
where $u$ and $v$ are any two scalar functions which satisfy the
homogeneous Helmholtz equation with parameter $k$ and $\hat\br$ is the
unit vector in the radial direction. It has been shown several
times\cite{papas,bouwkamp,wilcox} that knowledge of $E_r = \hat\br
\cdot \bE$ and $H_r = \hat\br \cdot \bH$ (in the volume) uniquely
determines all components of $\bE$, $\bH$. Furthermore, on a sphere,
the boundary value problem
\begin{equation}
\hat\br \cdot \bE|_{\partial B} = f, \qquad 
\hat\br \cdot \bH|_{\partial B} = g,
\end{equation}
is uniquely solvable if $f$ and $g$ are mean-zero
functions\cite{bouwkamp}.  Since both $E_r$ and $H_r$ satisfy the
homogeneous Helmholtz equation, they can be written in terms of a
spherical eigenfunction expansion, leading to the Mie series solution
for spherical scattering using the above Debye representation of the
electromagnetic fields. Briefly, in Mie scattering one usually
specifies $u$, $v$ (known as Debye potentials) using spherical partial
wave expansions:
\begin{equation}\label{eq-mierep}
\begin{aligned}
u(r,\theta,\varphi) &= \sum_{n,m} a_{nm} \, \psi_n^m(r,\theta,\varphi), \\
v(r,\theta,\varphi) &= \sum_{n,m} b_{nm} \, \psi_n^m(r,\theta,\varphi),
\end{aligned}
\end{equation}
where $\psi_n^m$ is the spherical partial wave function of degree $n$ and
order $m$,
\begin{equation}
\psi_n^m(r,\theta,\varphi) = h_n(kr) \, Y_n^m(\theta,\varphi),
\end{equation}
with $h_n$ the order $n$ spherical Hankel function of the first kind,
and $Y_n^m$ the spherical harmonic of degree $n$ and order $m$,
normalized so that $||Y_n^m||_2 = 1$. Using the coefficients
$a^\text{in}_{nm}$, $b^\text{in}_{nm}$ of a similarly expressed
incoming electromagnetic field, one can match modes depending on the
boundary conditions and analytically calculate $a_{nm}$,
$b_{nm}$\cite{papas,bouwkamp,wilcox,gimbutas-2013} for the scattered field.
The series for $u$, $v$ can be truncated depending on the number of
partial wave function needed to describe the incoming field.

We now extend the above spherical representation of Maxwell fields to
one which is immune from low-frequency breakdown and compatible with
planar geometries.  In Cartesian coordinates, it is easy to show
uniqueness of Maxwell fields which obey the Silver-M\"uller radiation
condition given $E_z = \bz \cdot \bE$ and $H_z = \bz \cdot \bH$. For
if $E_z=0$, $H_z=0$, the only electromagnetic fields satisfying
Maxwell's equations must be of the form:
\begin{equation}
\begin{aligned}
\bE &= a e^{ikz} \hat\bx + b e^{ikz} \hat\by, \\
\bH &= -b e^{ikz} \hat\bx + a  e^{ikz}  \hat\by,
\end{aligned}
\end{equation}
where $\hat\bx$, $\hat\by$ are unit vectors in the $x$ and $y$
directions, respectively, and $a$, $b$ are constants.  However, these
fields do not decay at infinity. Therefore, the coefficients $a$, $b$
must be set to zero in order for the field to obey the Silver-M\"uller
radiation condition.  Furthermore, if the electromagnetic fields are
generated by a finite collection of current, $\bJ$, located in a
volume $V$, then the $z$-components can be calculated easily using the
Lorenz gauge representation of fields, as in~(\ref{eq-lorenz}).  The
resulting expression is analogous to that derived by Bouwkamp and
Casimir\cite{bouwkamp} for the radial components, so we omit it here.
In order to extend their spherical result to the upper half-space, we
change representation~(\ref{eq-mierep}) slightly. It is clear that
another valid representation of Maxwell fields for $z>0$ is given by
\begin{equation}\label{eq-mierepz}
\begin{aligned}
\bE &= \nabla \times \nabla \times (u \, \bz) + i\omega\mu
\nabla \times (v \, \bz), \\ 
\bH &= \nabla \times \nabla \times
( v \, \bz ) - i\omega\epsilon \nabla \times ( u \, \bz),
\end{aligned}
\end{equation}
where the radial unit vector $\hat\br$ has been replaced with the unit
vector in the $z$-direction, $\bz$. It is easy to show a direct
correspondence between $u$, $v$ and the $z$-components of the
electromagnetic field:
\begin{equation}\label{eq-relation}
E_z(x,y,z) = -\left( \frac{\partial^2}{\partial x^2} + \frac{\partial^2}{\partial y^2}
 \right) u(x,y,z), \qquad H_z(x,y,z) = -\left( \frac{\partial^2}{\partial x^2} + \frac{\partial^2}{\partial y^2}
 \right) v(x,y,z) .
\end{equation}

In order to write $u$, $v$ in terms of the driving current $\bJ$, which
is now assumed to lie in a volume $V$ located in the lower half-space,
instead of expressing $u$, $v$ as partial wave expansions it is
convenient to form them in a manner which is compatible with the
planar geometry of the problem while still automatically satisfying
the Helmholtz equation.  For some $\widehat\sigma = \mathcal F[\sigma]$, 
$\widehat\tau = \mathcal F[\tau]$, we
assume that $u$, $v$ are generally of the form:
\begin{equation}\label{eq-uv}
\begin{aligned}
u(x,y,z) &= \int_{\bbR^2} g_k(x,y,z) \, \sigma(x,y) \, dxdy \\
&= \int_{\bbR^2} \widehat g_k(\xi,\eta,z) \, \widehat \sigma(\xi,\eta)
\, e^{i(\xi x + \eta y)} \, d\xi d\eta, \\
v(x,y,z) &= \int_{\bbR^2} \widehat g_k(\xi,\eta,z) \, \widehat \tau(\xi,\eta)
\, e^{i(\xi x + \eta y)} \, d\xi d\eta,
\end{aligned}
\end{equation}
where $\widehat g_k$ is the Fourier transform of the three-dimensional
Helmholtz Green's function, given in formula~(\ref{eq-som}). Using
this representation, $u$ and $v$ can be thought of as a superposition
of plane waves that obey the Sommerfeld radiation condition for the
Helmholtz equation.  Substituting the
spectral Sommerfeld formula for $g_k$ into the Lorenz gauge
representation of $\bH$ in~(\ref{eq-lorenz}), we see that
\begin{equation}
\begin{aligned}
\bH(\bx) &= \nabla \times \int_V \left( \frac{1}{8\pi^2}
\int_{\bbR^2} \frac{e^{-\sqrt{\xi^2+\eta^2-k^2}(z-z')}}
{\sqrt{\xi^2+\eta^2-k^2}} \, e^{i(\xi(x-x') + \eta(y-y'))} \, d\xi d\eta
\right) \bJ(\bx') \, dV' \\
&= \nabla \times \int_{\bbR^2} \widehat g_k(\xi,\eta,z) \left( 
\int_{V} e^{\sqrt{\xi^2+\eta^2-k^2}z'} \, \bJ(\bx') e^{-i(\xi x' + \eta y')}
dV' \right) e^{i(\xi x + \eta y)}  \, d\xi d\eta.
\end{aligned}
\end{equation}
This formula can be interpreted as constructing $\bH$ from a
superposition of attenuated transverse two-dimensional Fourier
transforms of the current $\bJ$.  The normal component $H_z$ is then
given by
\begin{equation}
H_z(\bx)
=  \int_{\bbR^2} \widehat g_k(\xi,\eta,z) \left( 
\int_{V} e^{\sqrt{\xi^2+\eta^2-k^2}z'} \, \left[ i\xi J_y(\bx')
- i\eta J_x(\bx') \right] e^{-i(\xi x' + \eta y')} \, 
dV' \right) e^{i(\xi x + \eta y)}  \, d\xi d\eta.
\end{equation}
Similarly, an expression for $\bE$ can be derived. We only
provide the formula for $E_z$, and not the derivation:
\begin{equation}
E_z(\bx) = \int_{\bbR^2} \widehat g_k(\xi,\eta,z) \left( \int_V
e^{\beta z'} \, 
\left[ i\omega\mu J_z(\bx') + \frac{\beta}{\epsilon}
  \, \rho(\bx') \right] \, e^{-i(\xi x' + \eta y')} \, dV' \right) \,
e^{i(\xi x + \eta y)} \, d\xi d\eta ,
\end{equation}
where $\beta = \sqrt{\xi^2+\eta^2-k^2}$ and we have used the
consistency condition $\nabla \cdot \bJ = i\omega \rho$.  Substituting
these expressions into relation~(\ref{eq-relation}), and using
representation~(\ref{eq-uv}) for $u$, $v$, we see that
$\widehat\sigma$ and $\widehat\tau$ are formally given by
\begin{equation}\label{eq-sigtau}
\begin{aligned}
\widehat\sigma(\xi,\eta) &= 
\frac{1}{\xi^2+\eta^2} \, \int_V e^{\sqrt{\xi^2+\eta^2-k^2} z'} \, 
\left[ i\omega\mu J_z(\bx') + \frac{\sqrt{\xi^2+\eta^2-k^2}}{\epsilon}
  \, \rho(\bx') \right] \, e^{-i(\xi x' + \eta y')} \, dV', \\
\widehat\tau(\xi,\eta) &= \frac{1}{\xi^2+\eta^2} 
\int_{V} e^{\sqrt{\xi^2+\eta^2-k^2}z'} \, \left[ i\xi J_y(\bx')
- i\eta J_x(\bx') \right] e^{-i(\xi x' + \eta y')} dV'.
\end{aligned}
\end{equation}
The integrands in the corresponding formulae for $\bE$, $\bH$ are
clearly integrable because of the extra $\xi$ and $\eta$ terms
introduced by differentiation. Therefore, equations~(\ref{eq-mierepz}),
(\ref{eq-uv}), and~(\ref{eq-sigtau}) provide a unique representation
of the electromagnetic field in the half-space $z>0$.

We can readily use the above method to calculate the solution of
scattering in $z>0$ from the perfect conducting half-space $z\leq 0$
using the boundary condition $\bn \times \bEtot = 0$ on $z=0$.  Using
representation~(\ref{eq-uv}) for the Debye potentials and
representation~(\ref{eq-mierepz}) for scattered fields $\bE$, $\bH$,
the components of $\bE$ on $z=0$ can be calculated as
\begin{equation}
\begin{aligned}
E_x(x,y,0) &= \int_{\bbR^2} \widehat g_k(\xi,\eta,0) \,
\left( i\xi \, \sqrt{\xi^2+\eta^2-k^2} \, 
\widehat\sigma(\xi,\eta)  - \omega \mu \eta \, \widehat\tau(\xi,\eta)
\right)  e^{i(\xi x + \eta y)} \, d\xi d\eta , \\
E_y(x,y,0) &= \int_{\bbR^2} \widehat g_k(\xi,\eta,0) \, 
\left( i\eta \, \sqrt{\xi^2+\eta^2-k^2} \, \widehat\sigma(\xi,\eta) +
\omega \mu \xi \, \widehat\tau(\xi,\eta) \right)
\, e^{i(\xi x + \eta y)} \, d\xi d\eta , \\
E_z(x,y,0) &= \int_{\bbR^2} \widehat g_k(\xi,\eta,0) \,
\left( \xi^2 + \eta^2 \right) \,  \widehat\sigma(\xi,\eta) \, 
e^{i(\xi x + \eta y)} \, d\xi d\eta.
\end{aligned}
\end{equation}
Enforcing the boundary condition $\bn \times \bEtot = 0$ yields
a two-by-two  linear system for $\widehat\sigma(\xi,\eta)$, 
$\widehat\tau(\xi,\eta)$:
\begin{equation}
\begin{aligned}
\widehat g_k(\xi,\eta,0) \left( i\xi \, \sqrt{\xi^2+\eta^2-k^2} \, 
\widehat\sigma(\xi,\eta) - 
\omega \mu \eta \,  \widehat\tau(\xi,\eta) \right)
  &= - \widehat E^\text{in}_x(\xi,\eta), \\
\widehat g_k(\xi,\eta,0) \left( i\eta \, \sqrt{\xi^2+\eta^2-k^2} \, 
\widehat\sigma(\xi,\eta) + \omega \mu \xi \,  
\widehat\tau(\xi,\eta) \right) &= - \widehat E^\text{in}_y(\xi,\eta).
\end{aligned}
\end{equation}
The functions $\widehat\sigma$ and $\widehat\tau$ are then formally
given as
\begin{equation}
\begin{aligned}
\widehat\sigma(\xi,\eta) &= \frac{8\pi^2 i}{\xi^2+\eta^2} \left( 
\xi \widehat E^\text{in}_x(\xi,\eta)) + \eta \widehat E^\text{in}_y(\xi,\eta)
 \right), \\
\widehat \tau(\xi,\eta) &= \frac{8\pi^2 \, \sqrt{\xi^2+\eta^2-k^2}}
{\omega\mu (\xi^2+\eta^2) } \left( 
\eta \widehat E^\text{in}_y(\xi,\eta)) - \xi \widehat 
E^\text{in}_x(\xi,\eta) \right),
\end{aligned}
\end{equation}
where we have explicitly substituted in the expression for $\widehat
g_k$.  It is interesting to point out that the terms $\xi \widehat
E^\text{in}_x + \eta \widehat E^\text{in}_y$ and $\eta \widehat
E^\text{in}_x - \xi \widehat E^\text{in}_y$ are proportional to the
Fourier transforms of the surface divergence and surface curl of the
incoming field $\bE^{\text{in}}$. This is not surprising considering
that non-physical currents used in the generalized Debye
representation are constructed from surface gradients and surface
curls on arbitrary smooth geometries; see the following section for a
brief introduction to this formulation.

If the incoming field $\bEin$, $\bHin$ is not known in terms of its
Fourier transform, but rather in terms of its Mie series or
component-wise partial wave expansion, then its Fourier
transform can be calculated analytically using the identities found in
the appendix of this note. Numerical schemes based on this observation
are currently being developed.

Generalizing the above approach to layered media geometries is
relatively straightforward, requiring several Fourier calculations and
matching of boundary conditions. It should be noted that no
low-frequency breakdown occurs in the resulting formulas for $\bE$,
$\bH$. This approach provides one alternative to the use of the dyadic
Green's function.  We skip the layered media calculation and instead
turn our attention to generalized Debye source methods in half-spaces
and layered media geometries.

\section{Generalized Debye sources}
\label{sec-debye}

The generalized Debye source representation of time-harmonic
electromagnetic waves\cite{EpGr,EpGrOn} is designed to provide a
unified framework for the representation of solutions to Maxwell's
equations in smooth geometries of arbitrary connectedness. The
resulting representations lead to well-conditioned, resonance-free
second-kind integral equations which are immune from low-frequency and
topological breakdown\cite{cools-2009,epstein-2013}.  In general, for
a simply-connected bounded scatterer $\Omega$ with boundary $\Gamma$,
the scattered fields $\bE$ and $\bH$ are constructed from mean-zero
scalar sources $r$, $q$ on $\Gamma$ using the fully symmetric
potential/anti-potential representation:
\begin{equation}\label{eq-ehdebye}
\begin{aligned}
\bE &= \sqrt{\mu} \left( ik \bA - \nabla \varphi - \nabla \times \bQ
\right), \\ \bH &= \sqrt{\epsilon} \left( ik \bQ - \nabla \psi + \nabla
\times \bA \right), \\
\end{aligned}
\end{equation}
where $\bA$, $\bQ$, $\varphi$, and $\psi$ are functions  defined by
the single-layer potentials
\begin{equation}
\begin{aligned}
\bA(\bx) &= \int_\Gamma g_k(\bx,\bx') \, \bj(\bx') \, da_{x'}, & \qquad 
\bQ(\bx) &= \int_\Gamma g_k(\bx,\bx') \, \bmm(\bx') \, da_{x'}, \\
\varphi(\bx) &= \int_\Gamma g_k(\bx,\bx') \, r(\bx') \, da_{x'}, & 
\psi(\bx) &= \int_\Gamma g_k(\bx,\bx') \, q(\bx') \, da_{x'}.
\end{aligned}
\end{equation}
From now on, the Helmholtz single-layer potential with kernel $g_k$ of
function $f$ will be denoted as $\mathcal S_k f$.  The functions $r$
and $q$ are known as {\em generalized Debye sources}.  In order for $\bE$
and $\bH$ in equation~(\ref{eq-ehdebye}) to satisfy Maxwell's
equations, the tangential vector fields $\bj$, $\bmm$ and scalar
functions $r$, $q$ must satisfy the consistency conditions
\begin{equation}
\nabla_\Gamma \cdot \bj = ikr, \qquad
\nabla_\Gamma \cdot \bmm = ikq,
\end{equation}
where $\nabla_\Gamma \cdot$ is the surface divergence.  Depending on
the boundary conditions, $\bj$ and $\bmm$ are explicitly constructed
from $r$ and $q$ such that that above consistency conditions are
automatically satisfied, as well as to ensure that the representation
is unique (i.e. no spurious resonances in the resulting integral
equations). For example, in the case where $\Omega$ is a
simply-connected, bounded perfect electric conductor ($\bn \times
\bE^\text{tot} = 0$ and $\bn \cdot \bH^\text{tot} = 0$) the tangential
fields $\bj$ and $\bmm$ are constructed as:
\begin{equation}\label{eq-jmpec}
\begin{aligned}
\bj &= ik\left( \nabla_\Gamma \triangle^{-1}_\Gamma r -
\bn \times \nabla_\Gamma \triangle^{-1}_\Gamma q \right), \\
\bmm &= \bn \times \bj,
\end{aligned}
\end{equation}
where $\nabla_\Gamma$ is the surface gradient and
$\triangle^{-1}_\Gamma$ is the inverse of the surface Laplacian
restricted to mean-zero functions.  When the boundary $\Gamma$ is
multiply connected, extra circulation conditions must be added to the
boundary conditions in order to determine the projection of $\bj$ and
$\bmm$ onto the space of harmonic vector fields along
$\Gamma$\cite{EpGr,EpGrOn}. We will skip this discussion here, as it
has been previously detailed in other papers by the author, and is
irrelevant in light of the unbounded planar geometries to be
addressed.  In order to determine $r$, $q$, we enforce two scalar
conditions on the surface of the conductor instead of one vector
condition. These scalar conditions are given by:
\begin{equation}
\mathcal S_0 \nabla_\Gamma \cdot \bn \times \bEtot = 0,
\qquad \text{and} \qquad \bn \cdot \bHtot = 0.
\end{equation}

Using these scalar boundary conditions, representations which
guarantee uniqueness and which lead to second-kind integral equations
for $r$ and $q$ that are free from low-frequency breakdown and
spurious resonances have been derived for the perfect electric
conductor\cite{EpGr} and the dielectric transmission
problem\cite{EpGrOn}. The generalized Debye source representation also
has the feature that $\bE$ and $\bH$ gracefully decouple as $\omega
\to 0$, leaving only the scalar potential terms in simply connected
geometries - $\bE$ depends only on $r$ and $\bH$ depends only on
$q$. This can be viewed as a generalization of the spherical
Lorenz-Mie-Debye representation.

\subsection{Generalized Debye sources on a perfectly
  conducting half-space}
\label{sec-gendebpec}

We will now derive a Sommerfeld-like formula for scattering from a
perfectly conducting half-space $z\leq0$ which is immune from
low-frequency breakdown, and decouples $\bE$ and $\bH$ for any value
of $\omega$.  This requires extending the generalized Debye source
representation to the domain $z>0$. In this geometry, the
representation simplifies in that the surface differential operators
reduce to their two-dimensional Cartesian counterparts, $\nabla$,
$\nabla \cdot$, and $\triangle$.  We first express the scalar
densities $r$ and $q$ in terms of their Fourier transform on the
$xy$-plane:
\begin{equation}
r(x,y) = \int_{\mathbb R^2} \widehat r(\xi,\eta) \, e^{i(x\xi + y\eta)}
\, d\xi d\eta, \qquad 
q(x,y) = \int_{\mathbb R^2} \widehat q(\xi,\eta) \, e^{i(x\xi + y\eta)}
\, d\xi d\eta,
\end{equation}
where it is implied that the mean-zero condition is satisfied via
$\widehat r(0,0) = \widehat q(0,0) = 0$.  In the case of the perfect
conductor, it now suffices to calculate the values of 
the two scalar boundary operators, $\mathcal S_0
\nabla \cdot \bE$ and $\bn \cdot \bH$, where from now on we will
abbreviate:
\begin{equation}
\mathcal S_0 \nabla \cdot \bE = - \mathcal S_0 \nabla \cdot \bn \times
\bn \times \bE,
\end{equation}
that is, we apply $\mathcal S_0 \nabla \cdot$ to the tangential
projection of $\bE$.  All operators in the generalized Debye source
representation are differential or convolutional; by writing the
Helmholtz equation's Green's function in its Sommerfeld representation
all operators diagonalize and can be applied via multiplication. We
recall from equation~(\ref{eq-som}) that the Fourier transform
$\widehat g_k$ of the Green's function for the Helmholtz equation
$g_k$ is given by:
\begin{equation}
\begin{aligned}
\widehat g_k(\xi,\eta,z) = \frac{1}{8\pi^2} 
         \frac{e^{-\sqrt{\xi^2+\eta^2-k^2}|z|}}{\sqrt{\xi^2+\eta^2-k^2}}.
\end{aligned}
\end{equation}
A single-layer potential of an integral function $f$, $\mathcal S_k f$,
can then be written as
\begin{equation}
\mathcal S_k[f](\bx) = \int_{\bbR^2} \widehat g_k(\xi,\eta,z) \,
\widehat f(\xi,\eta) \, d\xi d\eta.
\end{equation}
Evaluation of $\nabla$ and $\nabla \cdot$ are straightforward
multiplications by $i\xi$ and $i\eta$, and the inverse surface
Laplacian $\triangle^{-1}$ applied to a mean-zero function is given
by:
\begin{equation}
\triangle^{-1}[f](x,y) = \int_{\bbR^2} \frac{\widehat f(\xi,\eta)}
{-\xi^2-\eta^2} \, e^{i(\xi x + \eta y)} \, d\xi d\eta.
\end{equation}
After several tedious calculations, using the above identities and the
representations of $\bE$, $\bH$ from equation~(\ref{eq-ehdebye}), we
are able to calculate $\mathcal S_0 \nabla \cdot \bE$ and
$\bn \cdot \bH$ as
\begin{equation}
\begin{aligned}
  \mathcal S_0 \nabla \cdot \bE &= \sqrt{\mu} \int_{\bbR^2}
  \left( \xi^2 + \eta^2 -k^2  -ik \sqrt{\xi^2+\eta^2-k^2}  \right)
  \widehat g_0(\xi,\eta,0) \, \widehat g_k(\xi,\eta,0) \, 
   \widehat r(\xi,\eta) \, d\xi  d\eta, \\
\bn \cdot \bH &= \sqrt{\epsilon} \int_{\bbR^2}
  \left( \sqrt{\xi^2+\eta^2 - k^2} -ik \right)
  \, \widehat g_k(\xi,\eta,0) \, \widehat q(\xi,\eta) \, d\xi  d\eta.
\end{aligned}
\end{equation}
The Fourier transforms of $r$ and $q$ can then be calculated
as
\begin{equation}\label{eq-rqsol}
\begin{aligned}
\widehat r(\xi,\eta) &= -\frac{64\pi^4}{\sqrt{\mu}} 
\mathcal F\left[ \mathcal S_0 \nabla
  \cdot \bE^\text{in} \right] \frac{\sqrt{\xi^2 +
    \eta^2}} {\sqrt{\xi^2 + \eta^2 - k^2} - ik}, \\ 
\widehat q(\xi,\eta)  &= -\frac{8\pi^2}{\sqrt{\epsilon}}
\mathcal F\left[ \bn \cdot \bH^\text{in} \right]
 \frac{\sqrt{\xi^2 + \eta^2 - k^2}}
{\sqrt{\xi^2 + \eta^2 - k^2} - ik},
\end{aligned}
\end{equation}
where we have substituted in the expressions for the Green's functions
$\fg_k$ and $\fg_0$.  It is clear that in the previous formulas, the
equation for $r$ does {\em not} depend on the incoming magnetic field
$\bH^\text{in}$, and the equation for $q$ does {\em not} depend on the
incoming electric field $\bE^\text{in}$.  The unknowns have
effectively been decoupled, which is not surprising because of the
well-known separation of the vector wave equations for $\bE$ and $\bH$
into {\em transverse electric} and {\em transverse magnetic} fields.
In this respect, the generalized Debye sources are in one-to-one
correspondence with transverse electric and magnetic fields in the
presence of a perfectly conducting half-space. The expressions for
$\fr$ and $\fq$ in~(\ref{eq-rqsol}) are stable as $\omega \to 0$,
and when $\omega = 0$,
\begin{equation}\label{eq-rqsol0}
\begin{aligned}
\widehat r(\xi,\eta) &= -\frac{64\pi^4}{\sqrt{\mu}} 
\mathcal F\left[ \mathcal S_0 \nabla
  \cdot \bE^\text{in} \right] , \\ 
\widehat q(\xi,\eta)  &= -\frac{8\pi^2}{\sqrt{\epsilon}}
\mathcal F\left[ \bn \cdot \bH^\text{in} \right],
\end{aligned}
\end{equation}
analogous to the solution of a Neumann problem on a half-space.
We now derive similar Fourier-type solutions for scattering
in layered media using generalized Debye sources.

\subsection{Generalized Debye sources in layered media}

We now turn to the calculation of electromagnetic fields in layered
media using the generalized Debye source representation.  As in the
previous section, Section~\ref{sec-gendebpec}, we will derive formulae
for $\bE$, $\bH$ which are stable and decouple as the
frequency $\omega$ tends to $0$.  The simple two-layer case is
described first, with the generalization to $n$-layers later. We will
see that adjacent layers are coupled through the generalized Debye
sources; in the case of several layers, this coupling leads to a
banded system of linear equations whose bandwidth is {\em independent}
of the number of layers.

The calculations of the normal and tangential components of
generalized Debye source representations on the boundary of planar
layered media are similar to those in the previous section, but 
in order to ensure uniqueness we
require a different construction of the tangential vector fields $\bj$
and $\bmm$\cite{EpGrOn}. If the dielectric constants for $z>0$ are
$\epsilon_0$, $\mu_0$ and for $z<=0$ are $\epsilon_1$, $\mu_1$, then we
define the wavenumber $k_j = \omega\sqrt{\epsilon_j \mu_j}$.  The
fields $\bE_0$, $\bH_0$ and $\bE_1$, $\bH_1$ in the upper and
lower half-spaces, respectively, are given as
\begin{equation}
\begin{aligned}
  \bE_j &= \sqrt{\mu_j} \left( ik_j \mathcal S_{k_j}[\bj_j] -
    \nabla \mathcal S_{k_j}[r_j] - \nabla \times 
     \mathcal S_{k_j}[\bmm_j] \right), \\
  \bH_j &= \sqrt{\epsilon_j} \left( ik_j \mathcal S_{k_j}[\bmm_j] -
    \nabla \mathcal S_{k_j}[q_j] + \nabla \times 
     \mathcal S_{k_j}[\bj_j] \right),
\end{aligned}
\end{equation}
where the explicit layer potential dependence is shown instead of
using vector potential notation because of the variable wavenumber in
the Green's function. It is worth pointing out that if $z>0$, then
\begin{equation}
\mathcal S_k[f](\bx) = \frac{1}{8\pi^2} \int_{\bbR^2} 
\frac{e^{-\sqrt{\xi^2+\eta^2-k^2}z}}{\sqrt{\xi^2+\eta^2-k^2}}
\, \widehat f(\xi,\eta) \, d\xi d\eta,
\end{equation}
and if $z<0$,
\begin{equation}
\mathcal S_k[f](\bx) = \frac{1}{8\pi^2} \int_{\bbR^2} 
\frac{e^{\sqrt{\xi^2+\eta^2-k^2}z}}{\sqrt{\xi^2+\eta^2-k^2}}
\, \widehat f(\xi,\eta) \, d\xi d\eta.
\end{equation}
This is a direct consequence of the spectral formula for the Helmholtz
Green's function.  Sign mistakes in the exponent will lead to not only
incorrect formulas, but non-convergent integrands.  Furthermore, as
mentioned above, the tangential vector fields $\bj_j$, $\bmm_j$ are
constructed slightly differently\cite{EpGrOn} than in the perfectly
conducting case, with
\begin{equation}
\begin{aligned}
\bj_0 &= i\omega \left( \sqrt{\mu_0\epsilon_0} \nabla\triangle^{-1} r_0
- \epsilon_1 \sqrt{\frac{\mu_1}{\epsilon_0}} \bn \times \nabla \triangle^{-1}
r_1  \right),
& \qquad \bj_1 &= \bn \times \sqrt{\frac{\epsilon_0}{\epsilon_1}} \bj_0, \\
\bmm_0 &= i\omega \left( \sqrt{\mu_0\epsilon_0} \nabla \triangle^{-1} q_0
- \mu_1 \sqrt{\frac{\epsilon_1}{\mu_0}} \bn \times \nabla \triangle^{-1}
q_1 \right), &  
\bmm_1 &= \bn \times \sqrt{\frac{\mu_0}{\mu_1}} \bmm_0.
\end{aligned}
\end{equation}
The consistency conditions that are enforced automatically via the
construction of $\bj_j$, $\bmm_j$ are, as before,
\begin{equation}
\nabla \cdot \bj_j = ikr_j, \qquad
\nabla \cdot \bmm_j = ikq_j.
\end{equation}
As in the previous section, we now replace the generalized Debye
sources $r_j$, $q_j$ with their Fourier transform representations,
\begin{equation}
r_j(x,y) = \int_{\mathbb R^2} \widehat r_j(\xi,\eta) \, e^{i(x\xi + y\eta)}
\, d\xi d\eta, \qquad 
q_j(x,y) = \int_{\mathbb R^2} \widehat q_j(\xi,\eta) \, e^{i(x\xi + y\eta)}
\, d\xi d\eta.
\end{equation}
The Calder\'on preconditioned transmission boundary conditions to be
enforced on the interfaces are\cite{EpGrOn}:
\begin{equation}\label{eq-diebc}
\begin{aligned}[]
 \left[ \mathcal S_0 \nabla \cdot \bEtot \right] &=0, \qquad &
\left[ \bn \cdot \epsilon \bEtot \right] &= 0, \\
 \left[ \mathcal S_0 \nabla \cdot \bHtot \right] &=0,  &
\left[ \bn \cdot \mu \bHtot \right] &= 0.
\end{aligned}
\end{equation}
Using these representations, each of the operators can be written
in terms of its Fourier transform:
\begin{equation}\label{eq-lmeqs}
{\small
\begin{aligned}
\mathcal F\left[ \mathcal S_0 \nabla \cdot \bE_0
\right] &= \sqrt{\mu_0} \, \fg_{0}(\xi,\eta,0) \, 
\fg_{k_0}(\xi,\eta,0) \left( \left(\xi^2+\eta^2-k_0^2\right) \, \fr_0(\xi,\eta)
-i \omega \mu_1 \sqrt{\frac{\epsilon_1}{\mu_0}} \sqrt{\xi^2+\eta^2-k_0^2}
\, \fq_1(\xi,\eta) \right), \\
\mathcal F\left[ \mathcal S_0 \nabla \cdot \bE_1 \right] &= 
\sqrt{\mu_1} \, \fg_{0}(\xi,\eta,0) \, 
\fg_{k_1}(\xi,\eta,0) \left( \left(\xi^2+\eta^2-k_1^2\right) \, \fr_1(\xi,\eta)
+i \omega \mu_0 \sqrt{\frac{\epsilon_0}{\mu_1}} \sqrt{\xi^2+\eta^2-k_1^2}
\, \fq_0(\xi,\eta) \right),
\\
\mathcal F \left[ \mathcal S_0 \nabla \cdot \bH_0 \right] &= 
\sqrt{\epsilon_0} \, \fg_{0}(\xi,\eta,0) \, 
\fg_{k_0}(\xi,\eta,0) \left( \left(\xi^2+\eta^2-k_0^2\right) \, \fq_0(\xi,\eta)
-i \omega \epsilon_1 \sqrt{\frac{\mu_1}{\epsilon_0}} \sqrt{\xi^2+\eta^2-k_0^2}
\, \fr_1(\xi,\eta) \right), \\
\mathcal F \left[ \mathcal S_0 \nabla \cdot \bH_1 \right] &= 
\sqrt{\epsilon_1} \, \fg_{0}(\xi,\eta,0) \, 
\fg_{k_1}(\xi,\eta,0) \left( \left(\xi^2+\eta^2-k_1^2\right) \, \fq_1(\xi,\eta)
+i \omega \epsilon_0 \sqrt{\frac{\mu_0}{\epsilon_1}} \sqrt{\xi^2+\eta^2-k_1^2}
\, \fr_0(\xi,\eta) \right),
\\
\mathcal F\left[ \bn \cdot \epsilon_0 \bE_0 \right] &= 
\epsilon_0 \sqrt{\mu_0} \, \fg_{k_0}(\xi,\eta,0) \left( 
\sqrt{\xi^2 + \eta^2-k_0^2} \, \fr_0(\xi,\eta) +
i \omega \mu_1 \sqrt{\frac{\epsilon_1}{\mu_0}} \, \fq_1(\xi,\eta) \right), \\
\mathcal F \left[ \bn \cdot \epsilon_1 \bE_1 \right] &= 
-\epsilon_1 \sqrt{\mu_1} \, \fg_{k_1}(\xi,\eta,0) \left( 
\sqrt{\xi^2 + \eta^2-k_1^2} \, \fr_1(\xi,\eta) +
i \omega \mu_0 \sqrt{\frac{\epsilon_0}{\mu_1}} \, \fq_0(\xi,\eta)
\right), \\
\mathcal F\left[ \bn \cdot \mu_0 \bH_0 \right] &= 
\mu_0 \sqrt{\epsilon_0} \, \fg_{k_0}(\xi,\eta,0) \left( 
\sqrt{\xi^2 + \eta^2-k_0^2} \, \fq_0(\xi,\eta) -
i \omega \epsilon_1 \sqrt{\frac{\mu_1}{\epsilon_0}} \,
\fr_1(\xi,\eta) \right), \\
\mathcal F \left[ \bn \cdot \mu_1 \bH_1 \right] &= 
-\mu_1 \sqrt{\epsilon_1} \, \fg_{k_1}(\xi,\eta,0) \left( 
\sqrt{\xi^2 + \eta^2-k_1^2} \, \fq_1(\xi,\eta) -
i \omega \epsilon_0 \sqrt{\frac{\mu_0}{\epsilon_1}} \, 
\fr_0(\xi,\eta) \right),
\end{aligned} }%
\end{equation} 
where it is understood that $\bn = \bz$ {\em always}. This sign
convention needs to be especially consistent when dealing with
multiple layers. The boundary conditions in equation~(\ref{eq-diebc})
can now be applied through the above Fourier transforms, we omit the
expressions because their derivation is straightforward, albeit
somewhat lengthy. Instead, we give a condensed matrix version of
resulting system which displays the decoupling as $\omega \to 0$. In
matrix notation, in order to solve for $\fr_j$, $\fq_j$, we solve
\begin{equation}\label{eq-lmmat}
\left(
\begin{array}{cc}
A_{11} & \omega \, A_{12} \\
\omega \, A_{21} & A_{22}
\end{array}
 \right) \left(
\begin{array}{c}
\fr_0 \\
\fr_1 \\
\fq_0 \\
\fq_1 
\end{array}
\right)
=
\left( 
\begin{array}{c}
-\mathcal F\left[ \mathcal S_0 \nabla \cdot \bE^\text{in}\right] \\
-\mathcal F\left[ \bn \cdot \epsilon \bE^\text{in} \right] \\
-\mathcal F\left[ \mathcal S_0 \nabla \cdot \bH^\text{in}\right] \\
-\mathcal F\left[ \bn \cdot \mu \bH^\text{in} \right]
\end{array}
\right).
\end{equation}
For given values of $\xi$, $\eta$, the entries in $A_{ij}$ can be
easily derived from the formulas
in~(\ref{eq-lmeqs}). Examining~(\ref{eq-lmmat}), it is easy to see
that as $\omega \to 0$, the system becomes two-by-two block-diagonal,
with $\fr_0$, $\fr_1$ depending only on $\bE^{in}$ and $\fq_0$,
$\fq_1$ depending only on $\bH^{in}$. No numerical instabilities arise
in $A_{11}$ or $A_{22}$ as $\omega \to 0$, only a decoupling between
the $r_j$'s and the $q_j$'s.

\begin{figure}[t]
  \includegraphics[width=.6\linewidth]{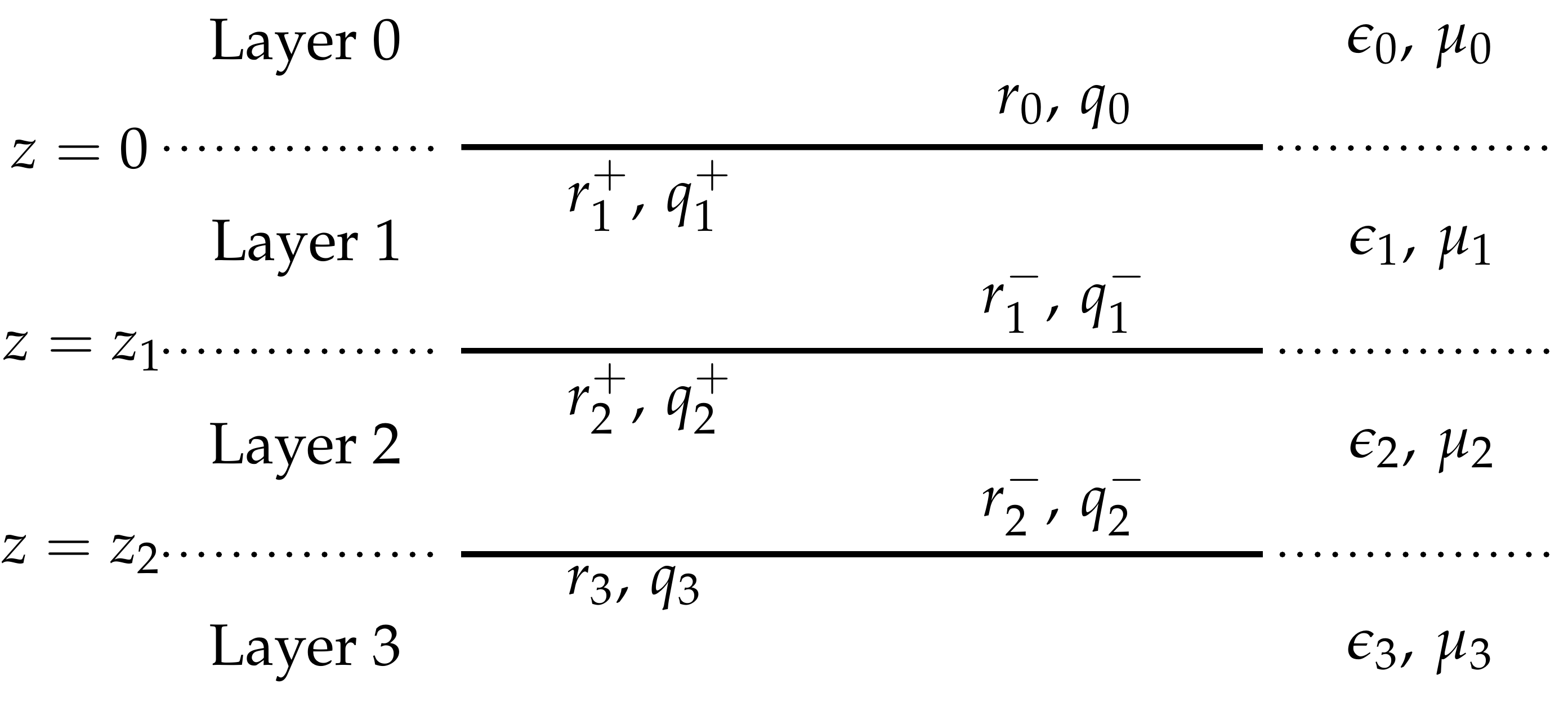}
  \caption{The distribution of unknowns $r_j$, $q_j$ in the multi-layer
    geometry.}
  \label{fig-multlayer}
\end{figure}

In the case of $n+1$ layers, for $j \neq 0,n$, the field in layer $j$
will be constructed from generalized Debye sources which are defined
on the boundary between layers $j-1$ and $j$ as well as on the
boundary between $j$ and $j+1$. For example, a single layer potential
in layer $j$ generated from a density on nearby interfaces would be
given by:
\begin{equation}
\mathcal S_k[f](\bx) = \frac{1}{8\pi^2} \int_{\bbR^2} 
\frac{e^{-\sqrt{\xi^2+\eta^2-k^2}(z-z_j)}}{\sqrt{\xi^2+\eta^2-k^2}}
\, \widehat f^-(\xi,\eta) \, d\xi d\eta +
\frac{1}{8\pi^2} \int_{\bbR^2} 
\frac{e^{\sqrt{\xi^2+\eta^2-k^2}(z-z_{j-1})}}{\sqrt{\xi^2+\eta^2-k^2}}
\, \widehat f^+(\xi,\eta) \, d\xi d\eta,
\end{equation}
where $z_{j-1}$ is the plane separating the $j^\text{th}$ and
$(j-1)^\text{th}$ layers, $z_j$ is the plane separating the
$j^\text{th}$ and $(j+1)^\text{th}$ layers, $f^+$ is the density
defined on $z=z_{j-1}$, and $f^-$ is the density defined on
$z=z_{j}$. See Figure~\ref{fig-multlayer} for a graphical
depiction. In this geometry, the corresponding formulas for $r_j$,
$q_j$ become slightly longer, but no more complicated than those
above.

\section{Conclusions}
\label{sec-conclusions}

In the preceding sections, we have presented two methods for the
solution to time-harmonic electromagnetic scattering problems in
planar geometries, namely half-spaces and layered media. Both methods
are immune from low-frequency breakdown, and rely on the Sommerfeld
representation of the Helmholtz Green's function, contrary to other
methods which focus on the construction of the physical dyadic domain
Green's function (using images).  The first of these representations,
in Section~\ref{sec-mie}, is based on the classic Lorenz-Mie-Debye
solution for electromagnetic fields in the exterior of a sphere. The
extension of this spherical representation to half-space Cartesian
geometries can be viewed as the natural limit of solutions to
Maxwell's equations in the exterior of an infinitely large sphere.

Secondly, in Section~\ref{sec-debye}, the generalized Debye source
representation of solutions to Maxwell's equations was extended to
both the half-space perfect conducting problem as well as the
bi-layered media scattering problem. In the presence of a perfectly
conducting half-space, the generalized Debye source approach
completely decouples the electric and magnetic fields into separate
equations which only involve half of the unknowns.  In the case of
layered media, there is a coupling of unknowns $r_j$, $q_j$ between
layers, as is expected. The bi-layer calculations can be immediately
extended to the multi-layer case, where only adjacent layers are
coupled.  This leads to a banded system of linear equations to solve,
whose bandwidth is independent of the number of layers involved.

As mentioned earlier, both new representations are based on the
spectral representation of the Green's function for the Helmholtz
equation. This approach allows the handling of arbitrary incoming
electromagnetic fields, not just dipoles or plane waves. The
convergence of such schemes relies on the decay rate of the transverse
Fourier transform of the incoming fields, which is exponential in the
distance of the driving current from the interface. For scatterers
located arbitrarily close to the layer-layer interfaces, image or
other analytical methods will be required in order to develop a fast
numerical scheme.  Such hybrid methods have already been developed in
the acoustic case\cite{oneil-imped}.

Our new extension of the generalized Debye source representation to
half-spaces and layered media retains all the desirable properties of
the analogous approach on bounded scatters - namely the absence of
low-frequency breakdown and a natural decoupling of the electric and
magnetic fields. As the frequency $\omega \to 0$, the generalized
Debye source equations simplify and the interaction between layers
decreases. This is in contrast to almost all methods based on the
electric dyadic Green's functions, which is inherently ill-scaled as
$\omega \to 0$.

Furthermore, formulas that convert spherical partial wave expansions
to their plane wave spectral representation have been provided.  These
formulas allow for a direct conversion between Sommerfeld
representations of fields and Mie series representations of fields.
This is analogous to the process by which spherical multipoles are
diagonally translated in analysis-based three-dimensional Helmholtz
fast multipole methods\cite{wideband3d}.

Future work on these methods will involve extending the generalized
Debye source approach to infinite layered media geometries which are
{\em not} purely planar, e.g. ones with ripples or localized
perturbations. Additionally, combining these methods with dielectric
or perfectly-conducting inclusions in the media (that cross
boundaries) will be necessary for practicality in industrial
applications. Fast and robust numerical algorithms based on both
approaches are currently being developed.

\begin{acknowledgements}
  The author's research was supported in part by the Air Force Office
  of Scientific Research under NSSEFF Program Award FA9550-10-1-0180.
\end{acknowledgements}

\appendix*
\section{Fourier transform of partial wave expansions}

In order to use Sommerfeld-like methods for the solution to scattering
problems, it is necessary to have access to the Fourier transform of
the incoming field along the scatterer.  The Fourier transform of the
incoming fields $\bEin$, $\bHin$ on the $xy$-plane does not need to be
calculated numerically if these fields are known in terms of their
component-wise partial wave expansions, i.e. as generated by a Mie
series. The Fourier representation (in cylindrical coordinates) of
outgoing partial wave functions is known
analytically\cite{nist,wideband3d,m+f} to be:
\begin{equation}\label{eq-pwexp}
\begin{aligned}
\psi_n^m(r,\theta,\varphi) &= h_n(kr) P_n^m(\theta) e^{im\varphi}
\\ &= \frac{(-i)^n i^m}{ik} \int_0^\infty
\frac{e^{-\sqrt{\lambda^2-k^2}z}}{\sqrt{\lambda^2 - k^2}} \, J_m
\left( \lambda r \right) \, e^{im\varphi} \, P_n^m\left(
\frac{i\sqrt{\lambda^2-k^2}}{k} \right) \, \lambda \, d\lambda,
\end{aligned}
\end{equation}
where $(r,\theta,\varphi)$ are the usual spherical coordinates, $(x,y,z)$
are Cartesian coordinates, and $z$ is assumed to be positive.  Such
representations are used for diagonal translation operators in fast
multipole methods for the three-dimensional Helmholtz
equation\cite{wideband3d}. An extra sign factor is required for $z<0$
to account for the parity of $P_n^m$.  Formula~(\ref{eq-pwexp})
can be derived via a calculation analogous to that in the proof of
Theorem 3.2 in Greengard and Huang\cite{greengard-huang}, or by
carefully applying the following differential
relation\cite{devaney-1974} to the Sommerfeld representation of $g_k$,
the Green's function for the Helmholtz equation:
\begin{equation}
  h_n(kr) Y_n^m(\theta,\varphi) = 
  c_n^m \left[ \left(\frac{1}{ik}\left[ \frac{\partial}{\partial x} +
     i \frac{\partial}{\partial y} \right] \right)^m P_n^{(m)}
    \left( \frac{1}{ik} \frac{\partial}{\partial z} \right) \right] 
\frac{e^{ik|\br|}}{k|\br|},
\end{equation}
where
\begin{equation}
\begin{gathered}
c_n^m = (-1)^m (-i)^m \sqrt{ \frac{(2n+1)(n-m)!}{4\pi (n+m)!}}, \\
P_n^{(m)} \left( \frac{1}{ik} \frac{\partial}{\partial z} \right) =
\frac{d^m}{du^m} P_n(u) |_{u = \frac{1}{ik} \frac{\partial }{\partial z}}.
\end{gathered}
\end{equation}
Here, the vector $\br = (x,y,z) = (r,\theta,\varphi)$ and $P_n$ is
the Legendre polynomial of degree $n$.

In short, if the Cartesian components of an incoming field $\bEin$,
$\bHin$ are known in terms of their partial wave expansions,
\begin{equation}
\begin{aligned}
\bEin(r,\theta,\varphi) &= \sum_{n,m} \left( a_{nm} \hat\bx +
b_{nm} \hat\by + c_{nm} \bz \right) \psi_n^m(r,\theta,\varphi), \\
\bHin(r,\theta,\varphi) &= \sum_{n,m} \left( e_{nm} \hat\bx +
e_{nm} \hat\by + f_{nm} \bz \right) \psi_n^m(r,\theta,\varphi),
\end{aligned}
\end{equation}
then the Fourier transform of each component can be calculated
on the $xy$-plane. For example, if
\begin{equation}
\begin{aligned}
E_x^\text{in}  &= \hat\bx \cdot \bEin \\
&= \sum_{n,m} a_{nm} \, \psi_n^m,
\end{aligned}
\end{equation}
then interchanging the sum and integration yields:
\begin{equation}
  E_x^\text{in}(x,y,z) = \frac{1}{ik} \int_0^\infty
  \frac{e^{-\sqrt{\lambda^2-k^2}z}}{\sqrt{\lambda^2 - k^2}} \,
  \sum_{n,m} a_{nm} \left( (-i)^n i^m \, J_m \left( \lambda\sqrt{x^2+y^2} \right) \,
    e^{im\varphi} \, P_n^m\left( \frac{i\sqrt{\lambda^2-k^2}}{k} \right)
  \right)\, \lambda \, d\lambda.
\end{equation}
This Fourier integral representation of $E^{in}_x$ can be efficiently
discretized\cite{wideband3d} in $\lambda$ along a contour which avoids the
singularity at $\lambda^2 = k^2$.


%

\end{document}